\documentclass[a4paper,11pt]{article}
\usepackage{pos}
\newcommand{\nuebar}{$\overline{\nu}_{e}$}
\title{Latest oscillation results from Daya Bay}

\author*[a]{Jinjing Li}

\affiliation[a]{Tsinghua University,\\
No.~30, Shuangqing Road, Haidian District, Beĳing, PRC}

\emailAdd{jinjing-li@mail.tsinghua.edu.cn}
\note{For the Daya Bay collaboration.}

\abstract{The Daya Bay reactor neutrino experiment is the first experiment that measured a non-zero value for the neutrino mixing angle $\theta_{13}$ in 2012. Antineutrinos from six 2.9 GW$_{\text{th}}$ reactors are detected in eight identically designed detectors deployed in two near and one far underground experimental halls. The near-far arrangement in km-scale baselines of anti-neutrino detectors allows for a high-precision test of the three-neutrino oscillation framework. Daya Bay's collection of physics data already ended on December 12, 2020. This proceeding shows the measurement results of $\theta_{13}$ and the mass-squared splitting $\Delta m^2_{32}$, based on the gadolinium-capture tagged sample in the complete data set with 3158 days of operation. The latest results on the hydrogen-capture-based oscillation analysis and search for light sterile neutrino are also summarized.}

\FullConference{The European Physical Society Conference on High Energy Physics (EPS-HEP2023)\\
 21-25 August 2023\\
Hamburg, Germany\\}

\begin{document}
\maketitle

\section{The Daya Bay Reactor Neutrino Experiment}
The Daya Bay Reactor Neutrino Experiment was located in Shenzhen, Guangdong, China. It started at the end of 2011 and aimed to detect electron anti-neutrinos (\nuebar) released from six commercial reactors with up to 2.9 GW$_{\text{th}}$ of thermal power. In order to observe neutrino oscillations at the kilometer-scale baseline to measure the then-unknown neutrino mixing angle $\theta_{13}$ precisely, the Daya Bay Experiment placed eight anti-neutrino detectors (AD) of identical design in the three experimental halls (EH). Two ADs were deployed in both EH1 and EH2, and they were $\sim$500 m away from the Daya Bay and Ling Ao reactors, which allows them to accurately measure the \nuebar~flux and energy spectrum from those reactors. Four ADs in EH3, which sites $\sim$1.6 km from the reactors, can observe the first maximal oscillation of the \nuebar~that is modulated by $\sin^2 2\theta_{13}$ shown in Eq.~\ref{eq:oscillation_prob}, 
\begin{equation}
\begin{aligned}\label{eq:oscillation_prob}
P_{\overline{\nu}_{e} \rightarrow \overline{\nu}_{e}} &= 1-\cos ^{4} \theta_{13} \sin ^{2} 2 \theta_{12} \sin ^{2} \Delta_{21} -\sin ^{2} 2 \theta_{13}\left(\cos ^{2} \theta_{12} \sin ^{2} \Delta_{31}+\sin ^{2} \theta_{12} \sin ^{2} \Delta_{32}\right)\\
  &=1-\cos ^4 \theta_{13} \sin ^2 2 \theta_{12} \sin ^2 \Delta_{21}-\sin ^2 2 \theta_{13} \sin ^2 \Delta_{e e}
\end{aligned}    
\end{equation}
where $\Delta_{ij} \equiv 1.267 \Delta m_{ij}^{2}~\left(\mathrm{eV}^{2}\right)[L~(\mathrm{m}) / E~(\mathrm{MeV})]$,   $\Delta m_{ij}^2$ is the mass-squared splitting between the mass eigenstates $\nu_i$ and $\nu_j$, the effective mass-squared difference $\Delta m^2_{ee}$ is related to the wavelength of the
oscillation observed at Daya Bay.

The multi-baseline measurement strategy relaxes the requirement of knowing the details of the fission process and operational conditions of the reactor. Therefore, it reduces the dominant systematics uncertainties, which are correlated among reactors and ADs. The identical design of each AD is the key to the effectiveness of uncertainty reduction. Reactor \nuebar~were detected via the inverse beta decay (IBD) reaction:
$\overline{\nu}_e+p\to e^++n$. The protons involved in the IBD reactions and contributed to the visible \nuebar~signal were largely provided by the liquid scintillator (LS). The main target for the oscillation analysis based on neutron capture on gadolinium ($n$Gd), namely the 20 tonnes of LS doped with $0.1\%$ gadolinium by weight (GdLS) was contained by the inner-most 3-m-diameter acrylic cylinder vessel (IAV). The IAV was enclosed by the outer 4-m-diameter acrylic vessel (OAV), which was filled with 22 tonnes of undoped LS. The LS was designed as the gamma catcher for the IBDs produced in GdLS, but later it also served as the main target for the oscillation analysis based on neutron capture on hydrogen ($n$H). The OAV was immersed in 36 tonnes of mineral oil (MO), which was contained by a 5-m-diameter stainless steel vessel (SSV) for shielding the outer radioactivity. Totally 192 8-inch diameter photomultiplier tubes (PMTs) were installed on the inner walls of the SSV to collect the \nuebar~signal produced in the GdLS and LS. The ADs in all halls were surrounded by two layers of water pools instrumented with PMTs that monitor cosmic rays and enable muon-related background reduction. More detailed descriptions of the experimental layout and detector structure can be found in Ref.~\cite{DayaBay:2015kir}.

Positron and neutron from the IBD reaction provided the remarkable event pair characteristics for the \nuebar. The positron deposits the energy quickly and gives the prompt signal. Neutron is dominantly captured on gadolinium or hydrogen and becomes the delayed signal. The details of the calibration and event reconstruction can be found in Ref.~\cite{DayaBay:2016ggj}.

Daya Bay's data collection started on December 24, 2011, and stopped on December 12, 2020. Due to the delay of the installation of the final two ADs in EH2 and EH3, and the early retirement of the first AD in EH1, there are a total of 217 days of 6-AD running period (from December 24, 2011, to July 28, 2012), 1524 days of 8-AD running period (from October 19, 2012 to December 20, 2016), and 1417 days of 7-AD running period (from January 26, 2017 to December 12, 2020).

\section{Neutrino Oscillation Results Based on Neutron Capture on Gadolinium}
Daya Bay's latest $n$Gd analysis utilizes a complete sample of $5.55\times10^6$ $n$Gd-IBD candidates acquired in 3158 days of operation~\cite{DayaBay:2022orm}. Details of the analysis process and techniques can be found in Refs.~\cite{DayaBay:2018yms,DayaBay:2016ggj}. Compared to the previous results~\cite{DayaBay:2018yms,DayaBay:2016ggj}, selection of IBD candidates has been optimized, energy calibration refined, and treatment of backgrounds further improved.

The non-uniformity and non-linearity of AD's energy response are both improved. The spallation neutron capture on Gd in the GdLS and delayed $\alpha$-particles from cascade decays of $^{214}\mathrm{Bi}$ and $^{214}\mathrm{Po}$ in the LS are used for the position-dependent correction. The new non-linearity correction curve is derived from the waveform output from a flash-ADC readout system running in parallel with the default ADC system of EH1-AD1 in 2016. The nonlinear response of the electronics is now calibrated for each channel. 

A new source of PMT flashers was observed in the 7-AD operation period that was not rejected by the previous criteria. An additional set of selection criteria is proposed targeting the characteristic charge pattern and temporal distribution of these new flashers. This new flasher criteria remove over $99\%$ of this instrumental background with an IBD selection efficiency of over $99.99\%$.

The $\beta$-n decay of cosmogenic radioisotopes $^9\mathrm{Li}/^8\mathrm{He}$ is the most significant background in this analysis. IBD candidates following the energetic muons are collected to evaluate their yield. To improve the identification of $^9\mathrm{Li}/^8\mathrm{He}$ from others, candidate events are separated into several samples based on the visible energy deposited by the muon in the AD and the distance between the prompt and delayed signals, $\Delta r$. The rates and energy spectra of the dominant cosmogenic radioisotopes are extracted with a simultaneous fit to 12 two-dimensional histograms defined by the different muon samples in the three halls for the two $\Delta r$ regions. We simply measure the sum of these two radioisotopes, due to their comparable lifetimes. This method provides higher statistics and better measurement of the low-energy part of the $\beta$ spectrum of $^9\mathrm{Li}/^8\mathrm{He}$ than the previous determination while reducing the rate uncertainty to less than $25\%$.

Due to the gradual failure of PMTs near the top of the water pools in the 7-AD period, some low-energy muons passed through the inner water pool (IWS) undetected. Therefore, a new muon-induced background, dubbed as ``muon-x'' became significant. This background typically consists of the muon as the prompt signal, and a Michel electron from muon decay or a product of muon capture or a spallation neutron as the delayed signal. The muon-x background was efficiently suppressed by rejecting events with a delayed signal less than 410 $\mu$s after a muon identified with an additional IWS PMT-hit multiplicity requirement:$[7, ~12]$ which led to a $<0.1\%$ loss in livetime. To determine the rate of this background together with fast neutrons, the prompt-energy spectra of the IBD-candidate sample are extended to 250 MeV and are fitted by the spectra of the fast-neutron sample and the IWS-tag muon-x sample with nHit = 7. The background rate is derived through extrapolation to the range of 0.7 MeV $<E_p <$12 MeV.

\begin{figure}[!htb]
    \centering
    \includegraphics[width=7cm]{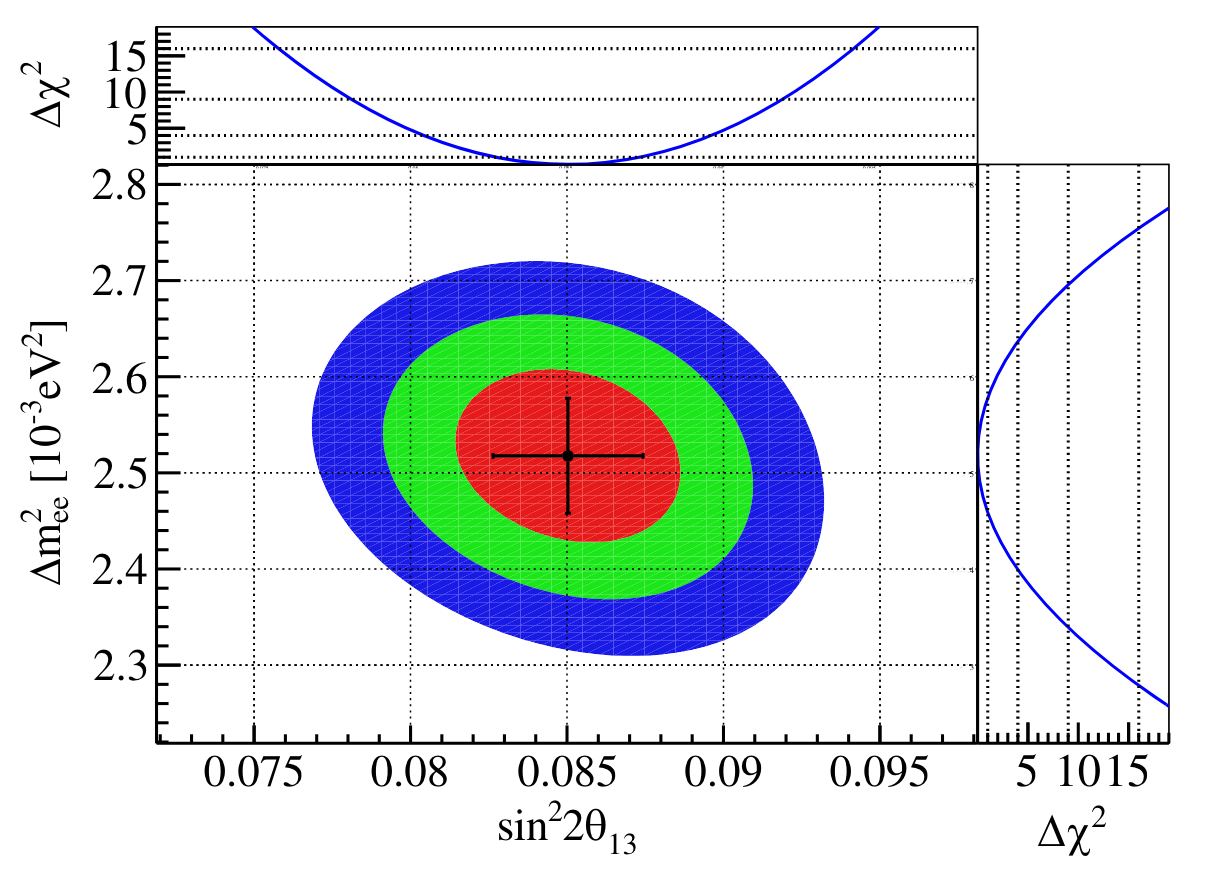}
    \includegraphics[width=7cm]{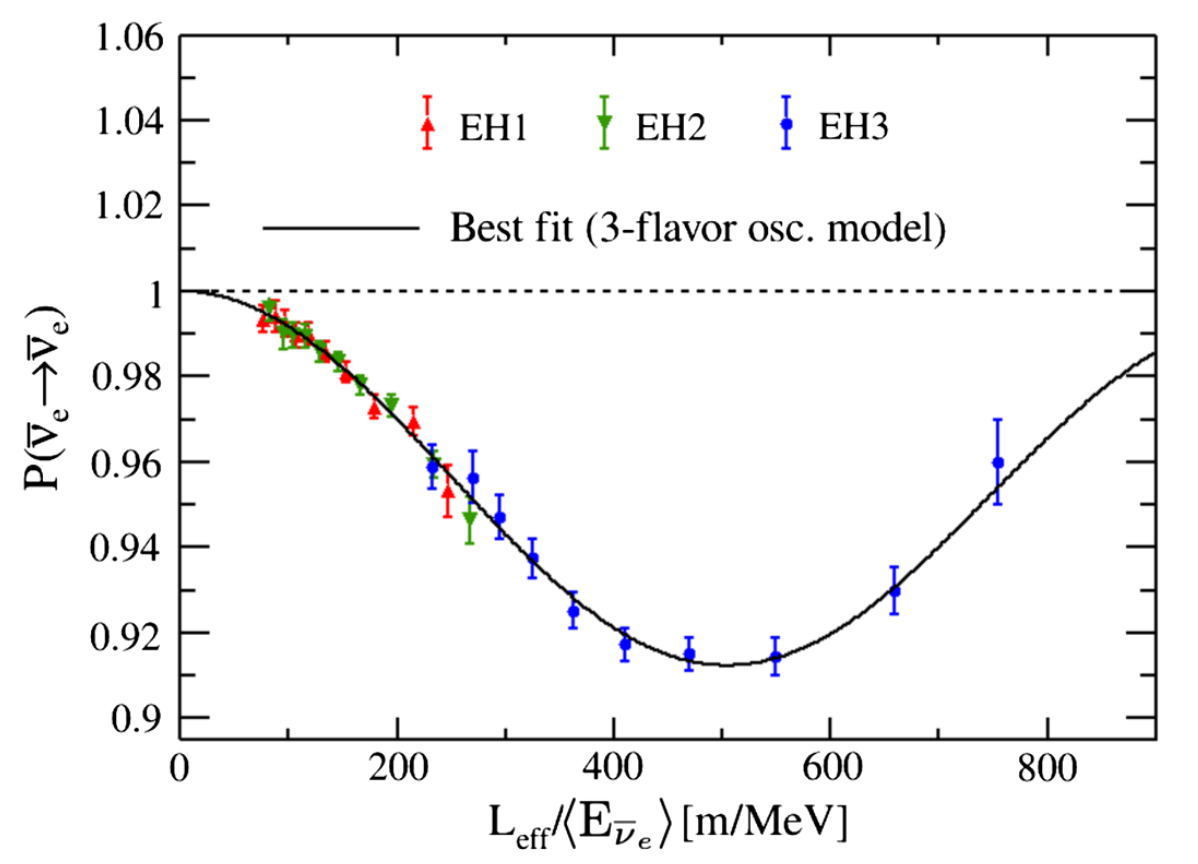}
    \caption{(Left) Error ellipses in the $\Delta m^2_{ee}$ vs.~$\sin^2 2\theta_{13}$ space. The point and error bars display the best-fit result and its one-dimensional 1-standard deviation. The colored contours correspond to 1, 2, and 3 standard deviations. (Right) Measured disappearance probability as a function of the ratio of the effective baseline $L_{\mathrm{eff}}$ to the mean antineutrino energy $\langle E_{\overline{\nu}_e} \rangle$.}
    \label{fig:nGd_fitting_contour}
\end{figure}

Fitting method B elaborated in Ref.~\cite{DayaBay:2016ggj} is used in this work to extract the oscillation parameters. Left plot  in Fig.~\ref{fig:nGd_fitting_contour} shows the allowed regions in the $\Delta m^2_{ee}$ vs.~$\sin^2 2\theta_{13}$ space. The best-fit gives $\sin^2 2\theta_{13} = 0.0851\pm0.0024$ with $\chi^2/\mathrm{NDF} = 559/517$, $\Delta m^2_{32} = (2.466\pm0.060)\times 10^{-3}~\mathrm{eV}^2$ for the normal mass ordering or $\Delta m^2_{32} = -(2.571 \pm 0.060) \times 10^{-3}~\mathrm{eV}^2$ for the inverted mass ordering. The normalized signal rate of the three halls as a function of $L_{\mathrm{eff}}/\langle E_{\overline{\nu}_e} \rangle$ with the best-fit curve superimposed is plotted in the right graph of Fig.~\ref{fig:nGd_fitting_contour}, where $L_{\mathrm{eff}}$ and $\langle E_{\overline{\nu}_e} \rangle$ are the effective baseline and average $\overline{\nu}_e$ energy, respectively.

\begin{figure}[!htb]
    \centering
    \includegraphics[width=15cm]{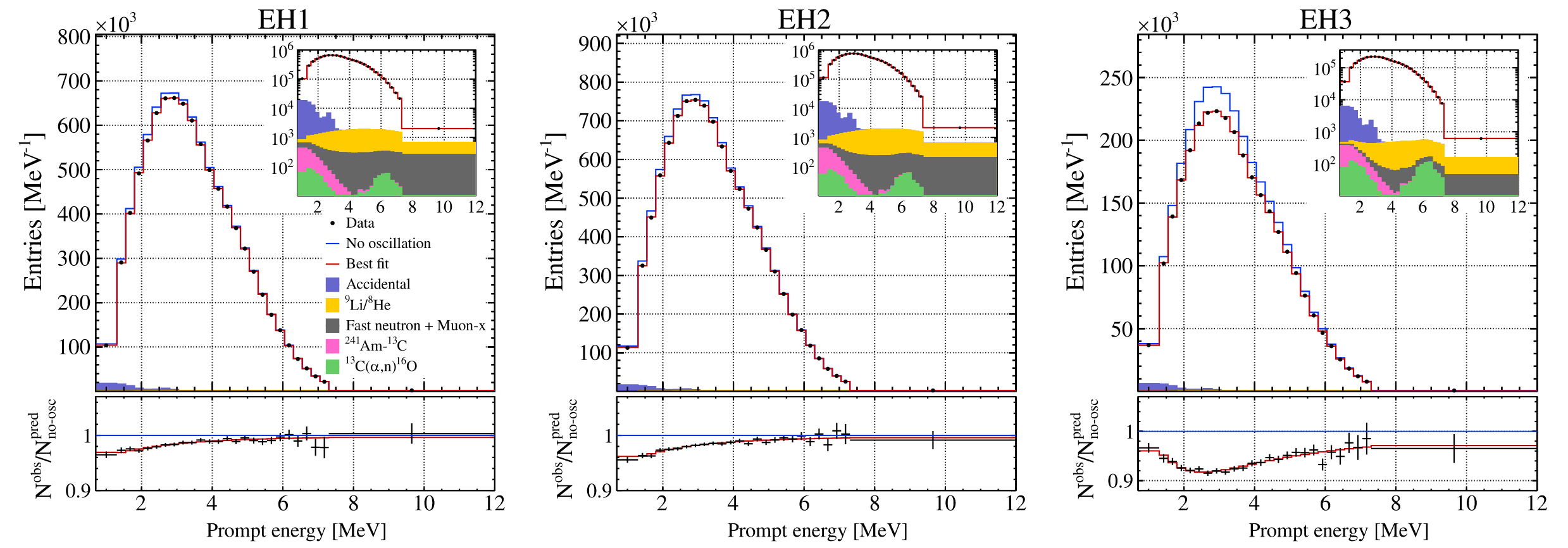}
    \caption{The measured prompt-energy spectra of EH1, EH2, and EH3 with the best-fit and no-oscillation curves superimposed in the upper panels. The backgrounds are apparent in the spectra with a logarithmic ordinate shown in the insets.}
    \label{fig:nGd_MvsP}
\end{figure}

As shown in Fig.~\ref{fig:nGd_MvsP}, the best-fit prompt-energy distribution agrees well with the observed spectra in each experimental hall.

\section{Neutrino Oscillation Results Based on Neutron Capture on Hydrogen}
In $n$Gd analysis, the \nuebar~sample is tagged by serval $\gamma$'s with a total energy of $\sim8$ MeV. This high energy labeling enables the $n$Gd analysis to get rid of most of the radioactive contamination. In $n$H analysis, the $\gamma$ generated by neutron capture by hydrogen is only 2.2 MeV. Therefore, the random pairing of low-energy radioactive background events constitutes an accidental coincidence background with event rates far exceeding $n$H-IBD. To reduce these backgrounds and bring the background rate to a level close to that of $n$H-IBD, we need to impose stricter selection criteria than $n$Gd analysis, which also introduces larger systematic uncertainty. Due to those significant differences, the results of $n$H and $n$Gd analysis are virtually independent. This allows $n$H analysis to provide a cross-validation for $n$Gd analysis results.

\begin{figure}[!htb]
    \centering
    \includegraphics[width=8.6cm]{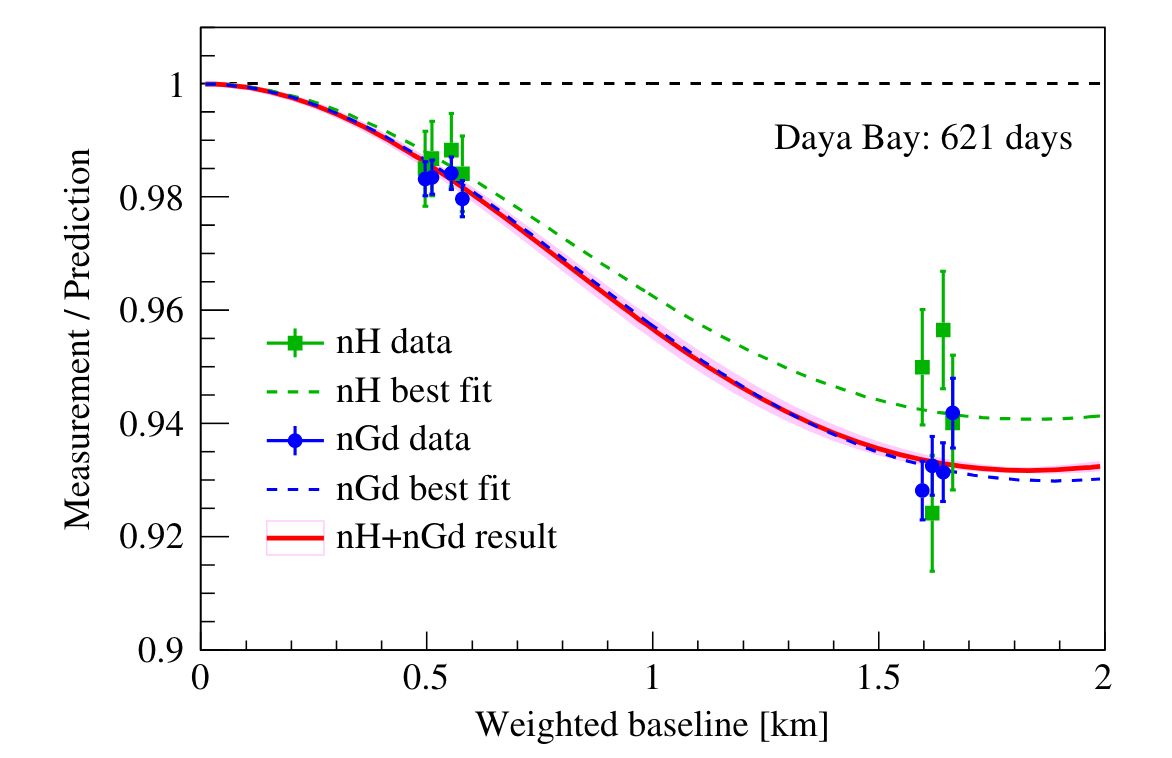}
    \includegraphics[width=6.4cm]{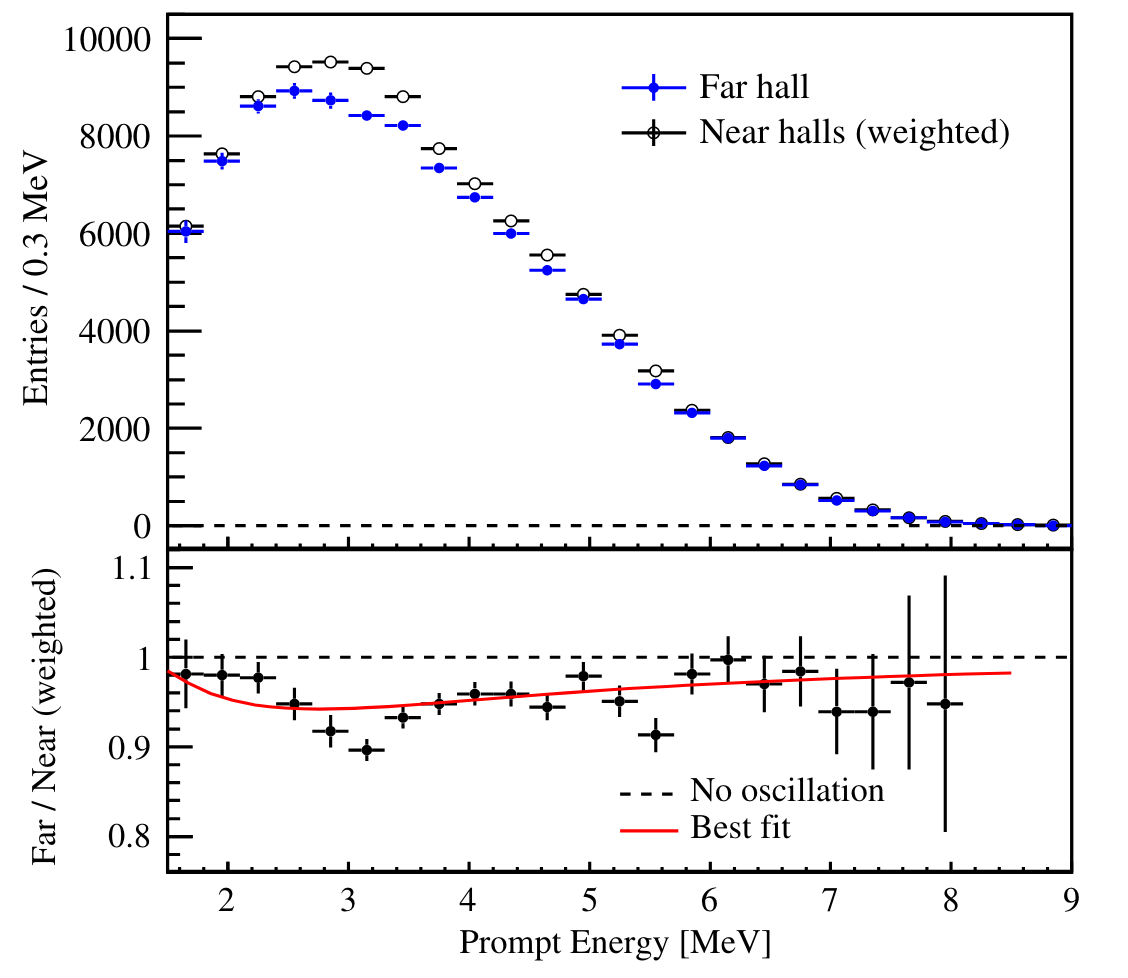}
    \caption{(Left) The ratio of the measured IBD rate to the non-oscillation prediction in each detector varies with the flux-weighted baseline. (Right) Measured prompt IBD spectra from far site and weighted near sites. The bin-by-bin ratio of the spectra is also shown together with the best-fit oscillation curve.}
    \label{fig:nH_fitting_contour}
\end{figure}

The latest result of $n$H analysis is still from 2016~\cite{DayaBay:2016ziq} with only 621 days of data. It gives the best-fit result: $\sin^22\theta_{13}=0.071\pm0.011$ with $\chi^2/\mathrm{NDF}=6.3/6$. The ratio of the measured IBD rate to the non-oscillation prediction for each AD vs.~flux-weighted baseline is shown in the left plot of Fig.~\ref{fig:nH_fitting_contour}. The prompt spectra of the far site are shown with that from the weighted near sites in the right plot of Fig.~\ref{fig:nH_fitting_contour}. Their ratio is shown together with the best-fit oscillation curve in the lower panel of the same plot, presenting the clear spectra distortion caused by the neutrino oscillation. Compared to this measurement with only rate deficit analysis, we have been putting a lot of effort into the $n$H analysis utilizing both the rate deficit and shape distortion with 1958 days of data set.
\section{Search for Light Sterile Neutrino}
\begin{figure}[!htb]
    \centering
    \includegraphics[width=7.5cm]{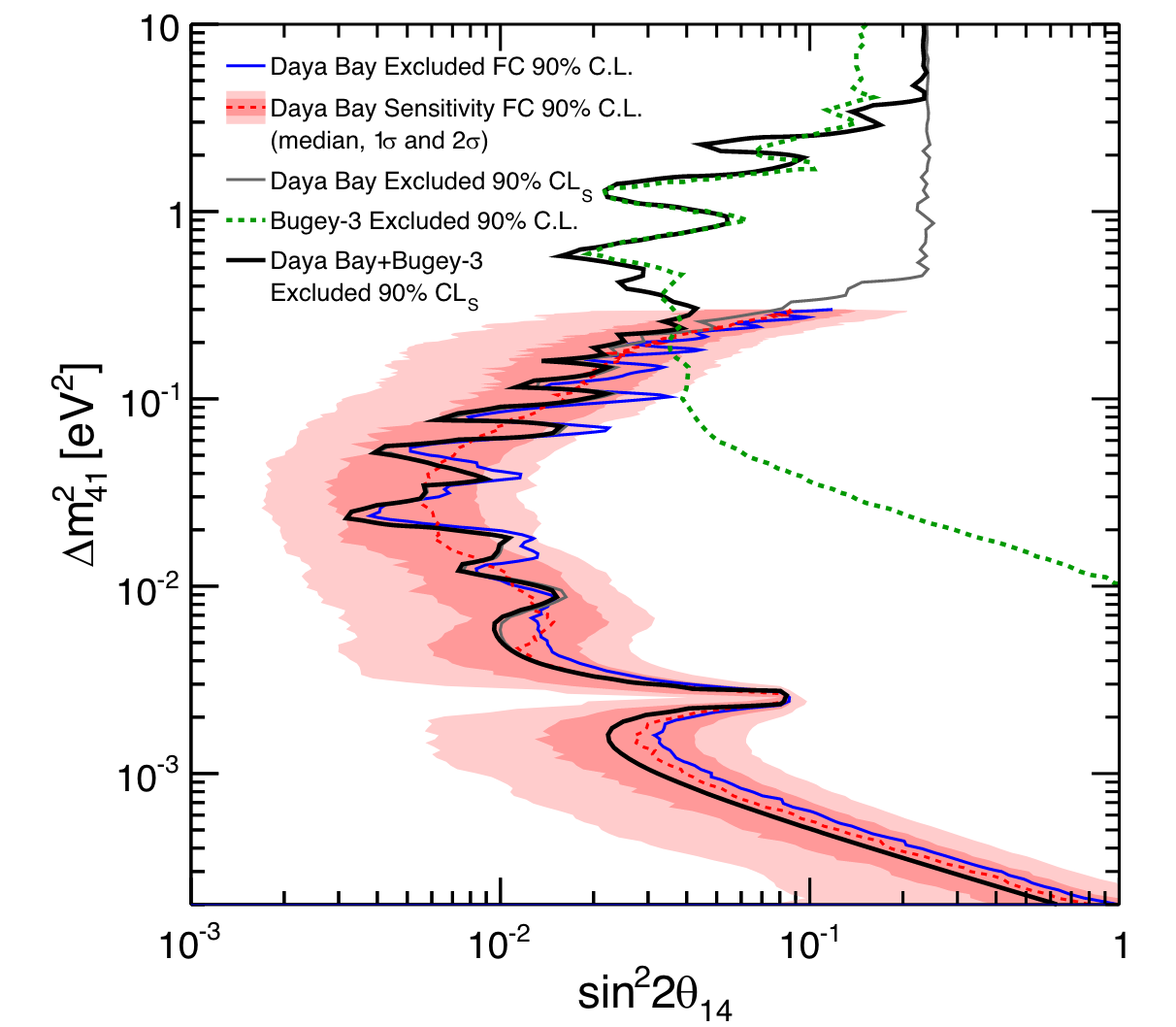}
    \includegraphics[width=7.1cm]{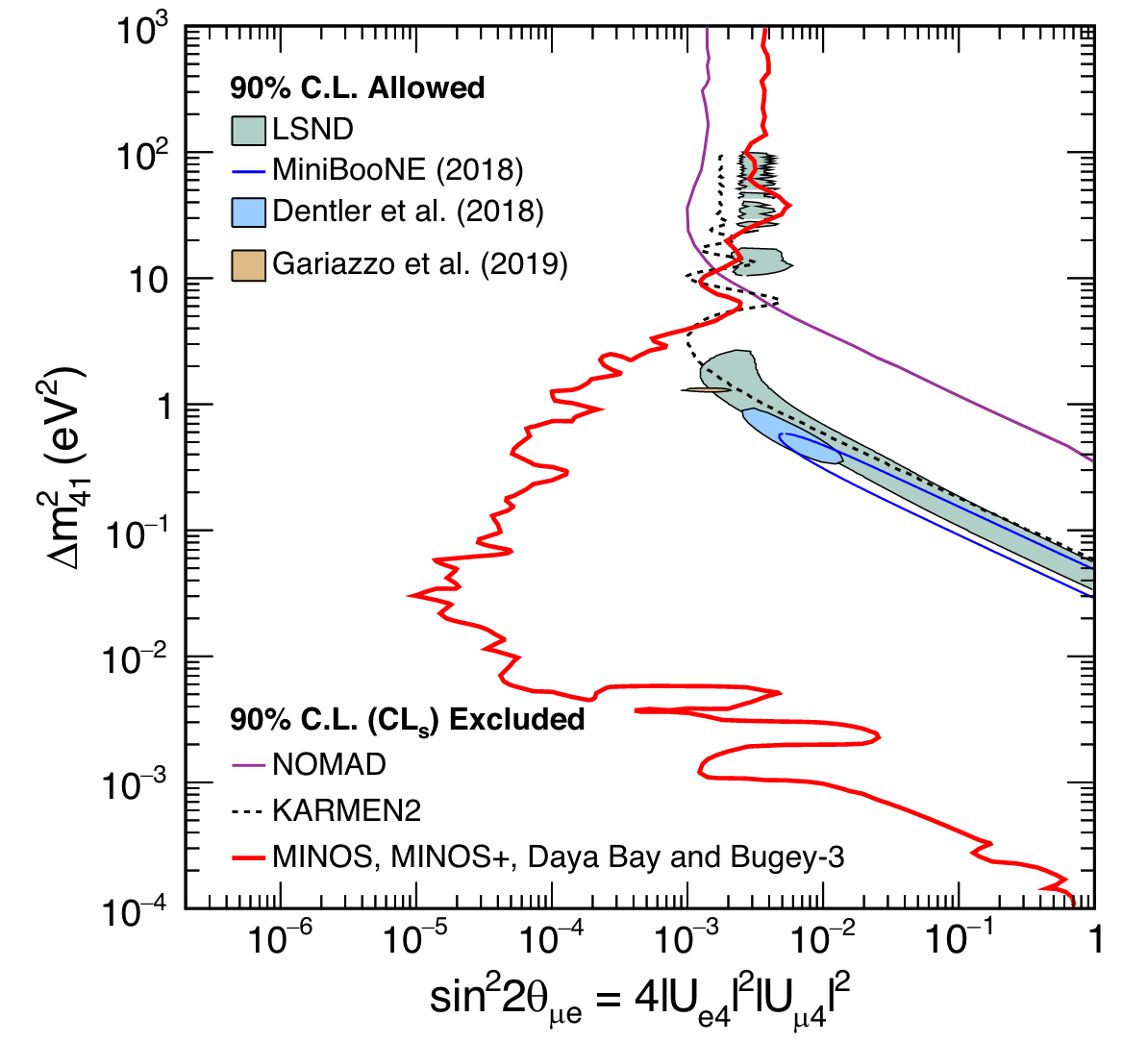}
    \caption{(Left) The exclusion and sensitive regions at $90\%$ C.L.~from the analysis of 1230 days of Daya Bay data. (Right) Comparison of the MINOS, MINOS+, Daya Bay, and Bugey-3 combined $90\%$ CLs limit on $\theta_{\mu e}$ to the LSND and MiniBooNE $90\%$ C.L.~allowed regions.}
    \label{fig:sterile}
\end{figure}

Most of the neutrino oscillation measurements with solar, atmospheric, reactor, and accelerator neutrinos can be fully explained with the three-neutrino-mixing framework. However, some anomalous measurements cannot be accommodated in this model, such as the excess of electron-like events in a $\nu_\mu$/$\overline{\nu}_\mu$ beam observed over short baselines by the LSND and MiniBooNE experiments, and reactor antineutrino anomaly observed over short baselines by reactor experiments.
The scenario of light sterile neutrino mixing is motivated by these anomalies, which could incorporate one sterile neutrino into the current framework with new parameters like $\sin^22\theta_{14}$ and $\Delta m^2_{41}$ ($\gg\Delta m^2_{32}$). To improve the constraints on this model from disappearance searches, the searches for \nuebar, $\nu_\mu$ and $\overline{\nu}_\mu$ disappearance in the Daya Bay and MINOS/MINOS+ experiments are combined~\cite{MINOS:2020iqj}, along with exclusion results from the Bugey-3 reactor experiment. The constraints on the $\theta_{\mu e}$ mixing angle are derived, which is over five orders of magnitude in the mass-squared splitting $\Delta m^2_{41}$, excluding the $90\%$ C.L.~sterile-neutrino parameter space allowed by the LSND and MiniBooNE observations at $90\%$ CLs for $\Delta m^2_{41}< 13~\mathrm{eV}^2$. Furthermore, the LSND and MiniBooNE $99\%$ C.L.~allowed regions are excluded at $99\%$ CLs for $\Delta m^2_{41}< 1.6~\mathrm{eV}^2$. The exclusion and sensitive regions using Daya Bay's data only and the combined results are shown in Fig.~\ref{fig:sterile}.

\section{Summary}
Daya Bay collaboration has reported the most precise measurement result of $\theta_{13}$ to date and one of the best measurements of $\Delta m^2_{32}$, using the complete data set with 3158 days of operation. Daya Bay has also provided the cross-validation for this $\theta_{13}$ measurement using the independent $n$H-IBD sample with 621 days of data. Improved constraints on the light sterile neutrino mixing and the combination with MINOS/MINOS+ and Bugey-3 experiments have also been presented. Using larger data sets, new results on the independent measurements of $\theta_{13}$ and $\Delta m^2_{32}$ from $n$H analysis and the constraint on light sterile neutrino mixing are expected.
\bibliographystyle{apsrev4-1}
\bibliography{ref}

\end{document}